# Scaling of the Maximum-Entropy Turbulence Energy Spectra

T.-W. Lee

Mechanical and Aerospace Engineering, SEMTE

Arizona State University, Tempe, AZ 85287

**Abstract-** The log-normal type of turbulence energy spectral function, derived from the maximum entropy principle, can be parameterized in terms of root turbulence variables including the Reynolds number. The spectral function is first compared with a number of experimental data sets, showing a very close agreement across the entire energy and length (wavenumber) scales. The peak wavenumber (m) and the width parameter ($C_2$) prescribe the spectral location and broadening when the Reynolds number increase, where $C_2$ has ~ $1/Re^m$ dependence. The energy magnitude is adjusted with a multiplicative factor. In this perspective, the inertial scaling from $k^{-3}$ to $k^{-5/3}$ when the flow transitions from two- to three-dimensions is explained as the increase in spectral width since the range of scales increases as $Re^{1/6}$ and $Re^{3/4}$ for two and three-dimensional turbulence, respectively. Energy spectra at various locations in channel flows are also reproduced using the same function, indicating applicability wherever local equilibrium is achieved. Therefore, based on a small number of scaling parameters the full energy spectra can be prescribed using the maximum-entropy formalism.

**INTRODUCTION**

The kinetic energy distribution in turbulence, often referred to as the power spectrum, is of importance for fundamental and practical reasons. Knowing the energy content across a range of length scales is useful in estimating the transport properties in the flow. The "spectral closure" has been an intensely studied in topic in fluid physics. Much effort has been expended on identifying the interaction mechanisms between the eddies at different scales. Some analytical methods have been developed since long ago, as in energy scaling in the inertial range ("$k^{-5/3}$-law" [1]), in two-dimensional turbulence [2], and a rather sophisticated method called EDQNM (eddy-damped quasi-normal Markovian) [3]. The references are intended as examples, among numerous others, and more complete reviews are available in the literature, e.g. Refs. 4 and 5. There have also been some attempts using the maximum entropy principle to derive the turbulence energy spectra [6, 7]. This is a sensible approach since turbulence consists of a statistically significant set of eddies which is expected to achieve rapid dynamical equilibrium. It is the Second Law (of thermodynamics) that dictates the state of this equilibrium, wherein the partition of energy is prescribed by the maximum entropy condition [8-10]. Using the maximum entropy principle along with the Lagrange multiplier method, the form of the energy distribution typically becomes an inverse exponential function of the constraint equations [10]. Then, the remaining step is to assert the constraints or boundary conditions of turbulence, to arrive at the energy distribution function [6, 7, 11].

There are several constraint conditions for turbulence that can be stipulated within the maximum entropy formalism. The most frequently used is the momentum conservation, written as some variations of the Navier-Stokes equation, such as Fourier-transformed [6] or vorticity [7]. There are more obvious and easily implementable boundary conditions, such as the limiting length scales and the energy conservation itself. I have shown that by using these as constraints within the maximum entropy principle the most probable energy distribution in turbulence has a lognormal form [11]. This functional form is intuitive and useful as it uses the basic turbulence variables to parameterize the energy distribution. Direct-interaction approximation [2] and EDQNM [3] tend to be complex in derivation and in the final form, reducing its accessibility. A compact, easily understandable theory is preferred, from aesthetic and application perspectives. In this work, I would like to demonstrate some scaling properties of the spectral form (Eq. 1 below) derived in a previous work [11], and demonstrate how it can be adapted in different turbulent flows.

**MAXIMUM-ENTROPY DISTRIBUTION FOR TURBULENCE ENERGY SPECTRA**

As we noted in our earlier work [11], the turbulence can be considered as a large ensemble of energetic eddies so that it must follow the Second Law or the maximum entropy principle: the final energy distribution is stipulated to be at the most stable maximum-entropy state, while obeying all the physical constraints. I have shown that the most probable distribution function under the specific constraints for turbulence has a lognormal form, Eq. 1. It follows from the application of the Lagrange multiplier method, which typically leads to an exponential decay function [10]. Insertion of the energy conservation constraint, along with dV=-1/k$^4$dk unit transform, leads to the following distribution function ("power spectra").

$$E(k) = \frac{C_1}{k^4} exp\{-C_2 u'^2 - C_3 k^2 u'^2\} \qquad (1)$$

The kinematic scaling for u' is m-log(k), heuristically determined from comparisons with data [11]. The full derivation is replicated in the Appendix, along with justification for this u'-scaling. $C_1$ and $C_2$ are the amplitude and spectral width parameters, respectively, while $C_3$ is the kinematic viscosity.

Let us first examine the efficacy of this lognormal form in reproducing the observed turbulence spectra in Figure 1, where we can see that the agreement between Eq. 1 and data is quite good for a wide range of Reynolds number. Moreover, the spectral coverage encompasses the entire energy and wavenumber range, starting from the energy-generating to Kolmogorov dissipation scale. At large Reynolds numbers ($Re_\lambda > 1000$), this means spans of about 10 orders of energy magnitude, and more than 5 for the wavenumber. There are no restrictions, theoretical or pragmatic, to be within the so-called "inertial range", as the maximum entropy principle produces the full energy state, across the entire wavenumber domain. At the lower energy-generating wavenumber, the distribution is truncated as the mode of turbulence production at these scales differ from one experiment to another, and it has not been input as a constraint. Thus, the final form of Eq. 1 may be referred to as a truncated (at $k_{min}$) lognormal-type distribution with a $k^{-4}$ modifier, in place of $k^{-1}$ in the conventional lognormal function. We also note that various $k^n$ type of scalings, the most prominent being n=-5/3 [1], are local or regional tangents to the lognormal distributions. For large Reynolds numbers, n=-5/3 tangent overlaps with the full E(k) over a good range of scales; however, it is a localized approximation. Unless the Reynolds number is very high the so-called inertial range is only a small portion of the full energy spectrum, with large sections near the energy-containing and dissipation scales missing from the picture.

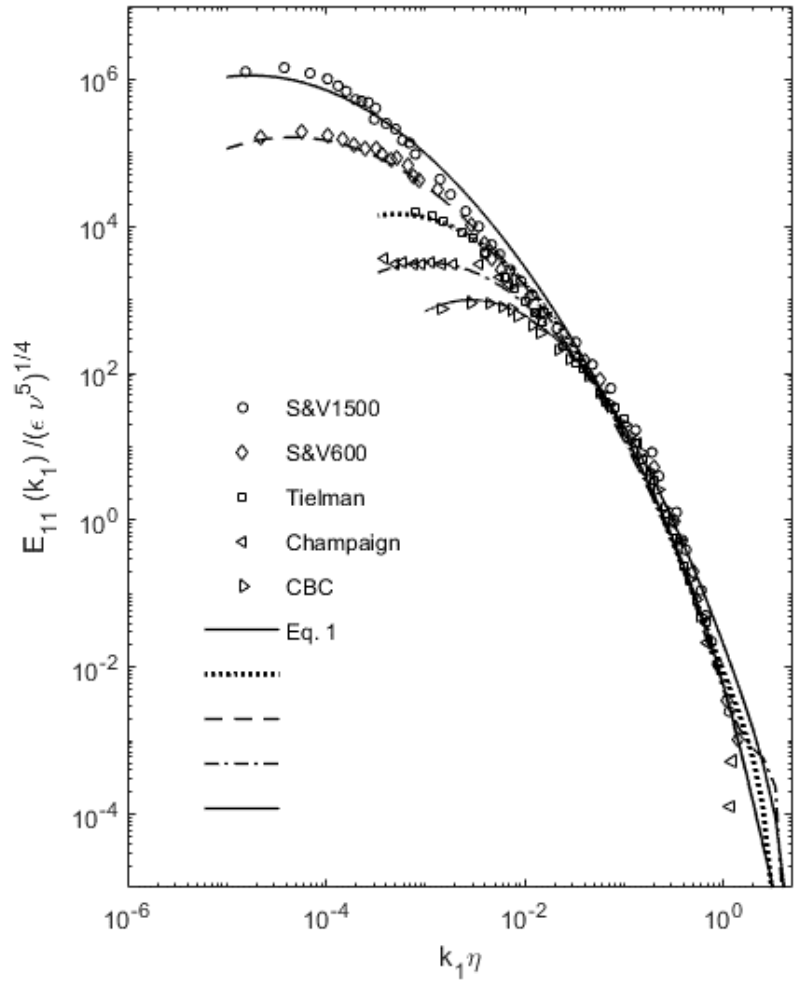

Fig. 1. Comparison of maximum-entropy spectra (Eq. 1) with experimental data [12-15]. For the data authors, see Table 1.

The lognormal form of Eq. 1 also makes it straight-forward to parameterize the distribution functions. For example, m is the logarithmic mean, which in this case corresponds to the $\log(k_m)$ where $k_m$ is the wavenumber at the peak energy scale, i.e. $1/k_m$ ~L is the energy-generating length scale. Given some variations in the energy generation processes, we will write $m \sim \log(k_m)$. $C_1$ is then the amplitude parameter, proportional to $E_{max}$, which in turn should scale with $\rho U^3/L$. However, given the various ways that the experimental conditions are described or sometimes obscured in the referenced works, we will take the simpler approach of $C_1 \sim E_{max}$. $C_2$ is the width parameter, inversely proportional to the logarithmic variance, $\sigma$. Because current form is not normalized, but instead vertically (energy magnitude) scaled by $C_1 \sim E_{max}$, $C_2$ has a large influence on the energy scale as well. Small $C_2$ corresponding to wide spectra also elevates the energy scale. Therefore at large Reynolds numbers, both the width and height of the energy spectra are increased by decreasing $C_2$. For this reason, $C_2$ is the key parameter in maximum-entropy turbulence energy distribution. Finally, $C_3 \sim \nu$ is the viscosity parameter, setting the maximum wavenumber and causing rapid dissipation close to this scale in Eq. 1. Thus, $k_m$, $E_{max}$, $Re_\lambda$, and $\nu$ furnish the parameters to prescribe the full turbulence energy spectra over the entire wavenumber space. The function parameters for the data in Figure 1 are listed in Table 1.

**Table 1.** Parameters of the energy spectra for various data sets.

| $Re_\lambda$ | $E_{max}/(\varepsilon \nu^5)^{1/4}$ | $(k\eta)_{@Emax}$ | $(k\eta)_{max}$ | C | m | A | Reference |
|---|---|---|---|---|---|---|---|
| 72 | 878 | 0.00439 | 1.164 | 0.315 | 0.5 | 0.0225 | CBC: Comte-Bellot and Corrsin [12] |
| 130 | 3183 | 0.001283 | 1.229 | 0.25 | 1.12 | 0.032 | Champaign [13] |
| 600 | 199000 | 0.0000578 | 1.429 | 0.16 | 2.5 | 0.05 | S & V: Saddoughi and Veeravalli [14] |
| 1282 | 15849 | 0.000777 | 0.685 | 0.25 | 0.45 | 0.01 | Tielman [15] |
| 1500 | 1446000 | 0.0000379 | 1.2 | 0.1475 | 2.5 | 0.05 | S & V: Saddoughi and Veeravalli [14] |

We can also examine more recent data by Kang et al. [16], where the evolution of the energy spectrum in decaying turbulence is experimentally observed. Eq. 1 follows this decay with a change in only the parameter $C_2$, with all other constants kept the same. The spectra tend to merge near the dissipation range, while diverging at the low wavenumbers as the Reynolds number decreases downstream. This spectral characteristic, and others, are reproduced by Eq. 1 with only a small variation in $C_2$ from 0.135 at $Re_\lambda = 716$ to 0.140 at $Re_\lambda = 626$. The current spectra (Eq. 1) overshoots the data in the $k_1 \sim 0.002$ to $0.05$ m$^{-1}$ range, although Gaussian-filtered data tend to smoothen this segment [16]. Straight or triangular (in log-log plots) probability distributions are rarely observed, as energy distributions in nature tend to be of exponential decay (Maxwell) or lognormal-type (Planck distribution), depending on the physical constraints [16]. During a single experiment, it is also difficult to cover larger spans of the Reynolds number. Nonetheless, the lognormal-type distribution (Eq. 1) exhibits parametric variations that mimic the Reynolds number dependence.

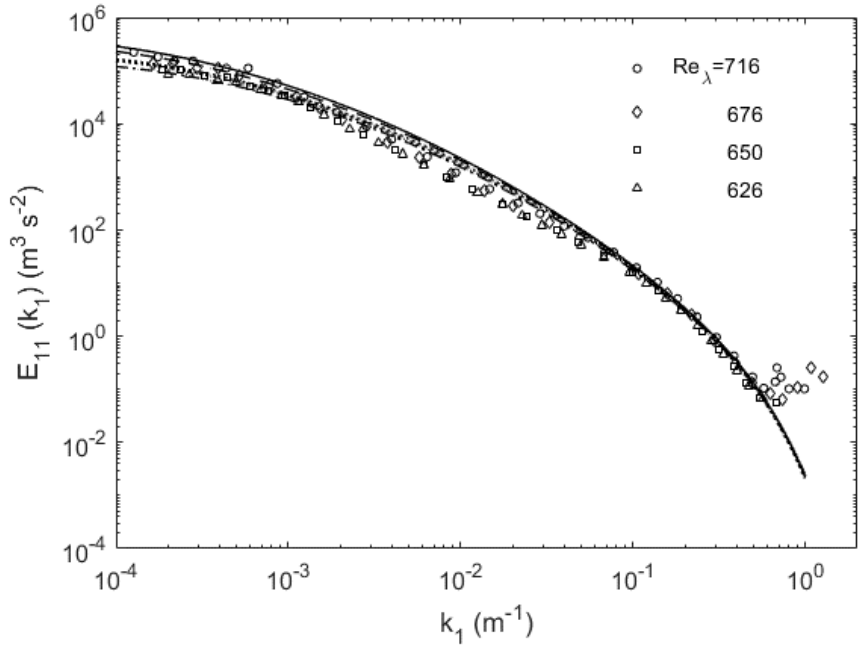

**Fig. 2. Comparison of the maximum-entropy spectra with experimental data of Kang et al. [16]. $C_2$ only was varied from 0.135 at $Re_\lambda$ = 716 to 0.140 at $Re_\lambda$ = 626, with $C_1$, m, and ν fixed in Eq. 1.**

The lognormal behavior of turbulence energy spectra is also evident in inhomogeneous flows, such as channel flows, as shown in Fig. 5, in spite of the spatial transport and non-local energy cascades. Power spectra, taken at various points in the channel, all follow lognormal form to a remarkable degree, when compared with DNS (direct numerical simulation) data [17] for $Re_\lambda$ = 180, 395 and 590. Current maximum-entropy distribution replicates the observed spectra quite well, except near the wall in the mid-wavenumber range ($k_x/k_{max}$ ~ $10^{-1}$) where Eq. 1 undershoots the data. Also, at low Reynolds number ($Re_\lambda$ = 180), there is a small discrepancy at low wavenumbers. Nonetheless, the overall reconstruction of the energy spectra using Eq. 1 is quite good. It appears that local equilibrium is achieved at high Reynolds numbers and the state of maximum entropy exists, so that lognormal energy spectra is applicable to inhomogeneous flows. This extends the applicability of current concept to inhomogeneous flows at sufficiently high Reynolds numbers. Again, lognormal behavior is prevalent across nearly the entire range of scales, far beyond the so-called inertial range.

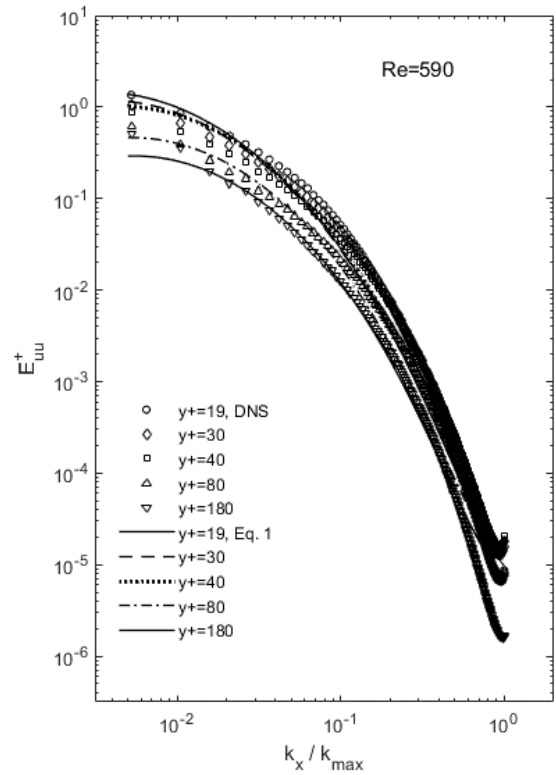

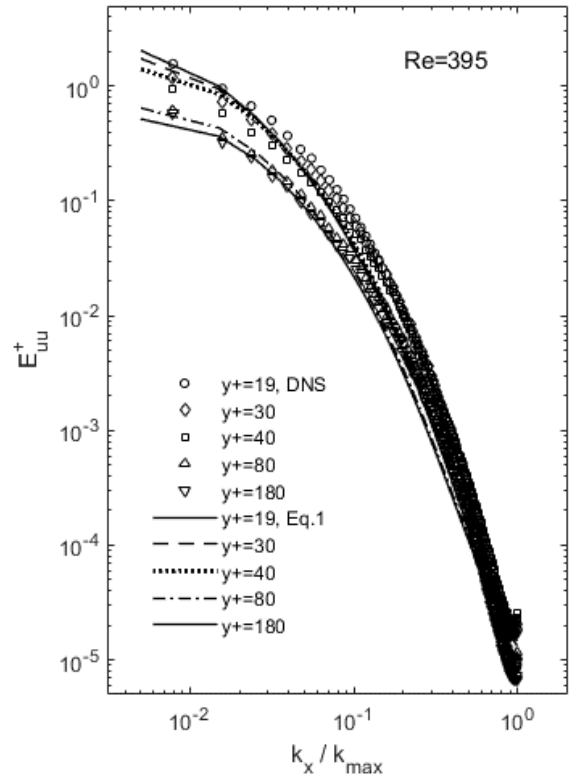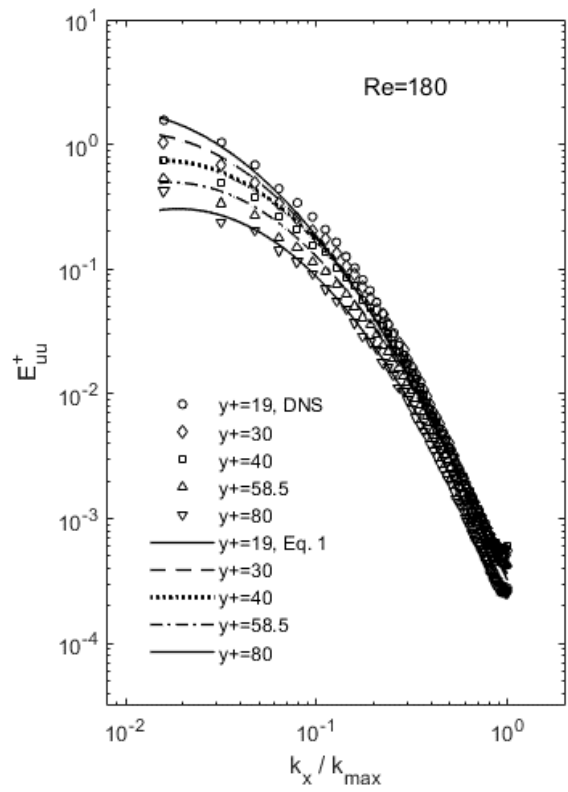

**Figure 3.** Turbulence energy spectra at various distances (y+) from the wall for channel flows, for $Re_\lambda$ = 590 (top), 395 and 180 (bottom). Lines are from Eq. 1, compared with DNS data [17].

The discussions above point to the validity of the log-normal form of energy spectra in turbulence. Since it has been derived from the fundamental Second Law (of thermodynamics), it has universal applicability in globally or locally equilibriated turbulence. It also explains the observed $k^{-m}$ scaling exponents which tended to vary across different experiments. In particular, m goes from -3 to -5/3 in the inertial range when the flow transitions from two- to three dimensions. This change in the slope occurs when the width of the spectra is broadened ($C_2$ decreases in Eq. 1). For the same Reynolds number, the length scale range increases for three-dimensional turbulence and therefore we expect a decrease in the parameter $C_2$. Figure 2 shows that reducing $C_2$ from 0.1 to 0.034 replicates the transition of energy spectra from two to three dimensions, as compared with large-scale atmospheric data [18]. For meridinal and potential temperature spectra, there were similar transitions when $C_2$ = 0.1 → 0.05 and 0.1625 → 0.05, respectively. The dimensional transition at these scales involve a substantial change in $C_2$, a decrease by 1/20 to 1/30, from two- to three-dimensional fields.

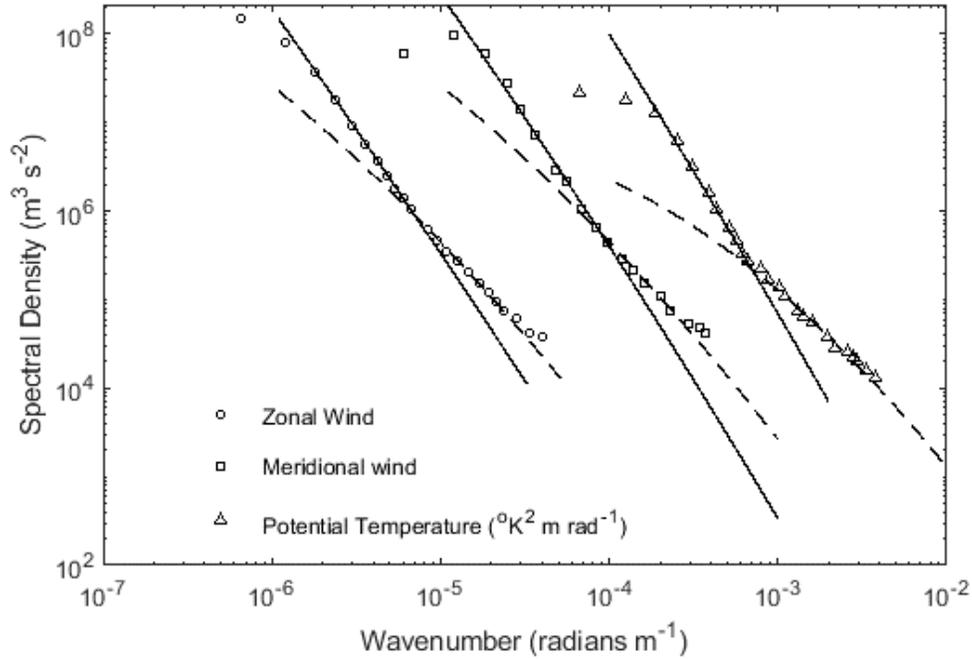

**Figure 4. Transition of the energy spectra from two- (solid lines) to three-dimensions (dashed). Data (symbols) are Tung and Orlando [18].**

**SCALING OF THE TURBULENCE ENERGY DISTRIBUTION**

It would be interesting to see the scaling behavior of the energy spectra observed above, in the context of the most probable distribution form (Eq. 1). In Table 1, we have columnized the spectral parameters associated with the data shown in Fig. 1, for which the $Re_\lambda$ spans from 72 to 1500. The lognormal distribution parameters, $C_1$, $C_2$ and m are included. The magnitude and length scale of the energy generation vary from one experiment to another, and thus the main parameter that determines the Reynolds number dependence is $C_2$, which shows monotonic decrease with increasing Reynolds numbers in Table 1, except for the data by Tielman [15]. Even though the Reynolds number variation in Kang's data [16] is just from 626 to 716, it does bear some consistency within the same experimental settings. Thus, we can plot $C_2$ as a function of the Reynolds number for all these data, in Fig. 5. $C_2$ goes as $Re_\lambda^{-1/5}$, after properly scaling the wavenumber by the Kolmogorov length scale. The outlier point for the Tielman's data possibly requires a corrected scaling, except for which there appears to be a monotonic dependence for $C_2$ on the Reynolds number.

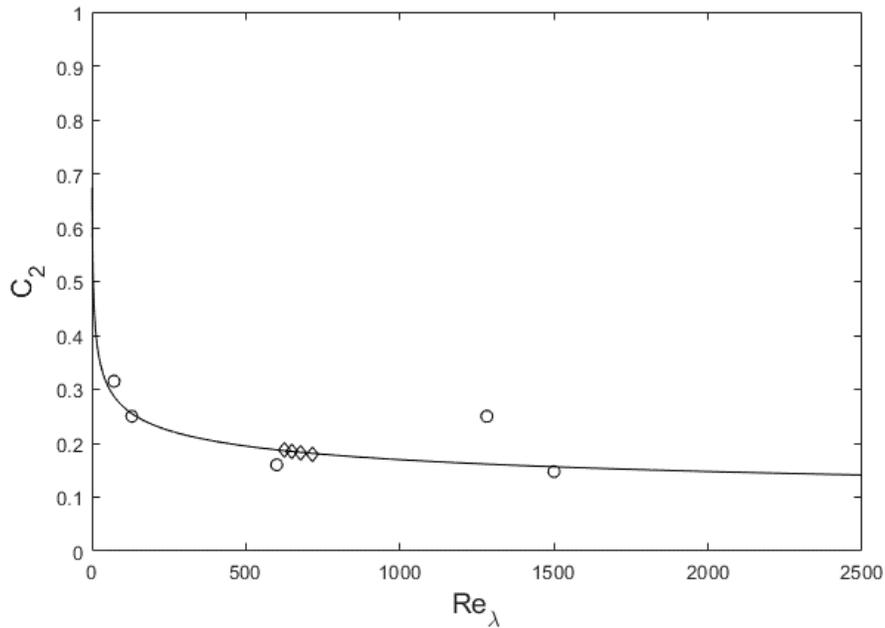

**Figure 5. Reynolds number scaling ($\sim Re_\lambda^{-1/5}$) of the spectral parameter, $C_2$.**

For the channel flow spectra (Figure 3), we can similarly analyze the function parameters, m and $C_2$, in Figures 6 and 7, respectively. We recall that the spectra vary spatially and also at different Reynolds numbers, so that the graphs are plotted as a function of y+ for $Re_\lambda$ = 180, 395 and 590. Since the local $u'^2$ is the driver for the energy spectra, they are plotted in an alternate (right-side) vertical axis. The wavenumber, m, at which the energy is predominantly produced increases slightly but consistently with both $u'^2$ and the Reynolds number. This is one possible difference from the homogeneous flow: m decreases or the length scale of energetic eddies increases at high Reynolds numbers. In channel flows, the integral scale decreases relative to the flow scale (channel width) with the local Reynolds number and u', so that m is weakly but proportional to u' (m increases with local u' and the global Reynolds number). This leaves the parameter, $C_2$, again as the key determinant of the energy spectra. Consistent with the current picture, $C_2$ is inverted with respect to both the $u'^2$ and the Reynolds number, again representative of the spectral broadening. $C_1$ is a measure of the spectral magnitude, and it follows the $u'^2$ variations in space (y+). Note that the peak $u'^2$ is relatively independent of the Reynolds number when normalized by the friction velocity, $u_\tau^2$, as shown in Figure 8. $C_1$ actually decreases with increasing Reynolds number, the reason being that any minute decrease in $C_2$ amplifies the spectral intensity much faster than $C_1$.

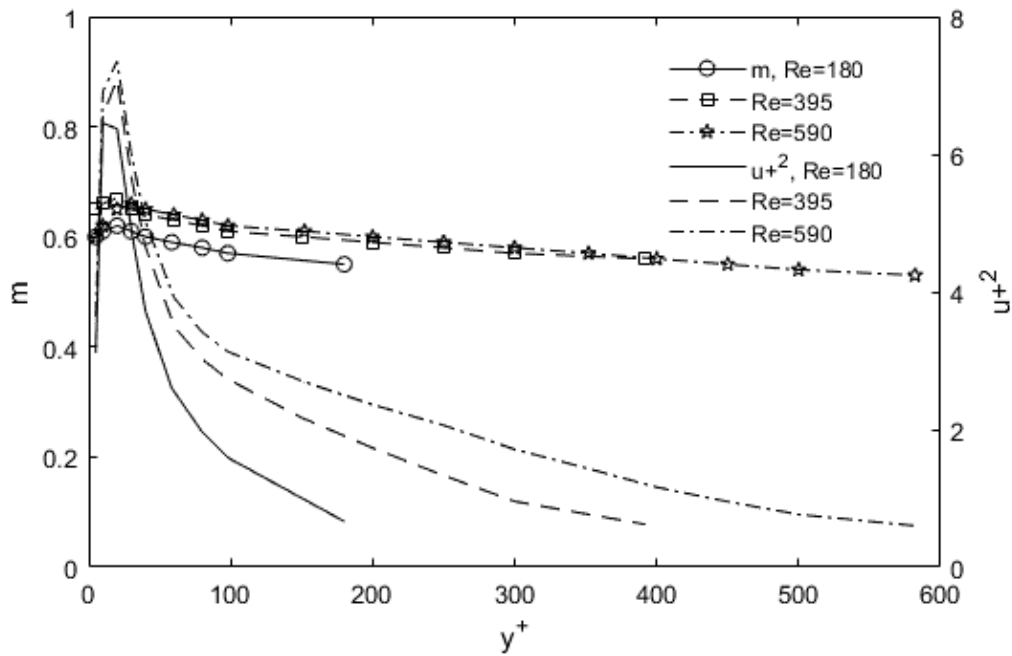

**Figure 6. Scaling of the spectral parameter, m, for channel flows.**

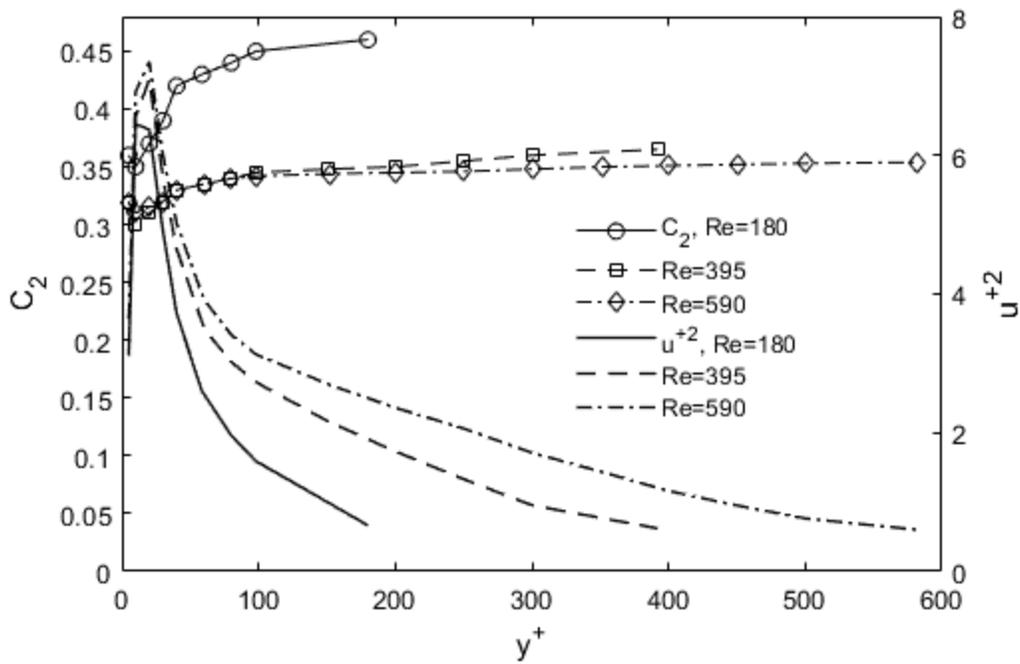

**Figure 7.** Scaling of the spectral parameter, $C_2$, for channel flows.

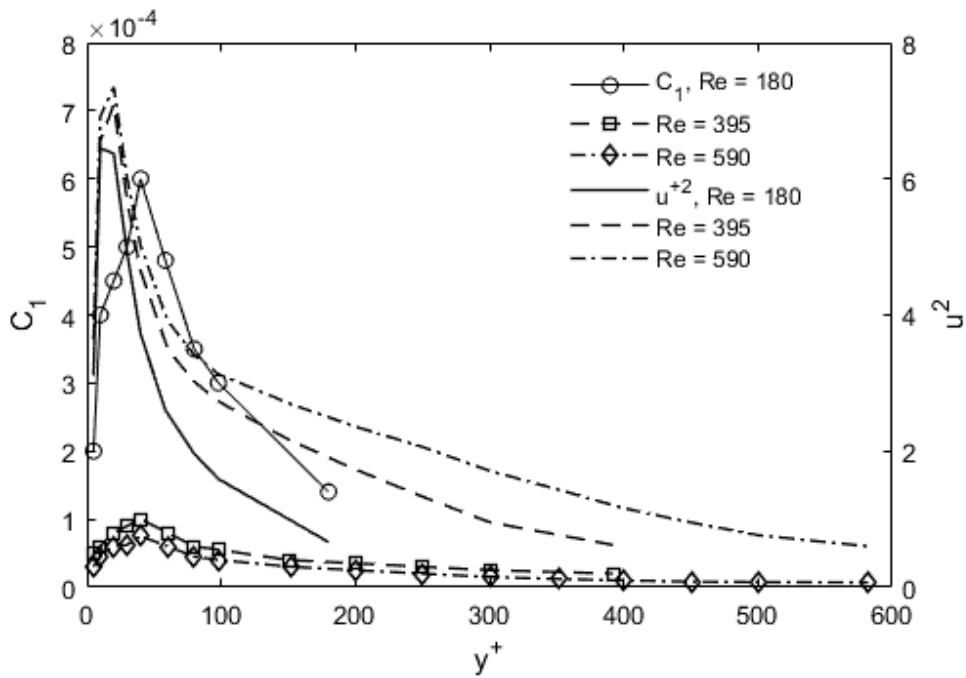

**Figure 8.** Scaling of the spectral parameter, $C_1$, for channel flows.

We can summarize the parametric scaling of the energy spectra, by varying one of the variables while fixing the other. Figure 9 shows that the peak energy scale increases by nearly three orders of magnitude for a modest change in $C_2$, from 0.5 to 1.5. This is accompanied by the spectral broadening effect of this parameter. The wavenumber parameter, m, shifts the spectra, while also affecting the magnitude, since it is in the exponential function. Finally, the viscosity, $C_3$, only modifies the spectral form near the dissipation range, and controls the rate of descent due to rapidly depleting kinetic energy.

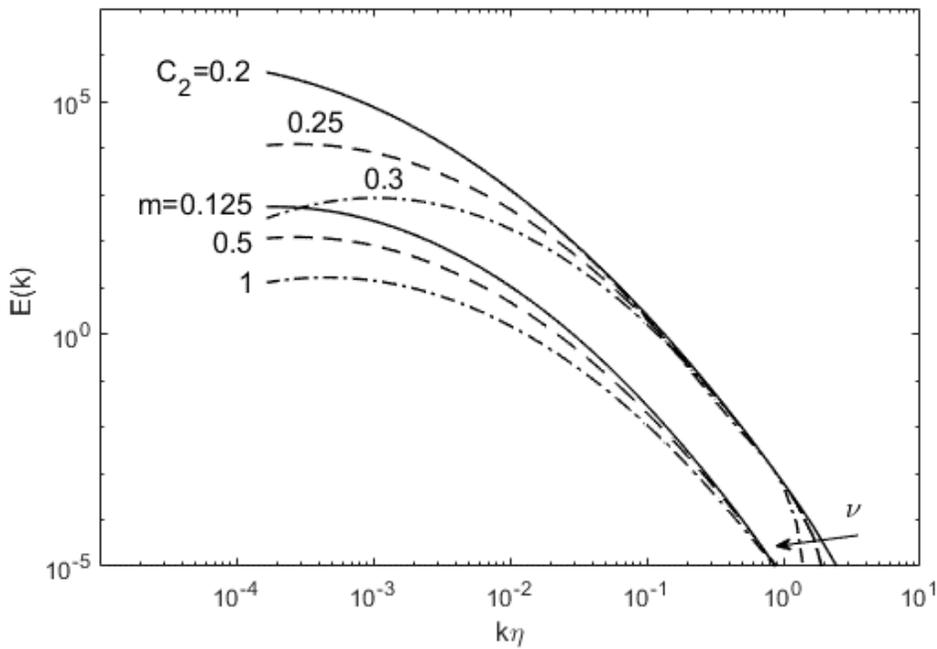

Figure 9. Parametric variations of $C_2$, m, and $\nu$.

**CONCLUSIONS**

Previous methods to derive the turbulence energy distribution have involved some detailed scale-to-scale transport modeling or approximations. Spectral closure models such as DIA (direct interaction approximation) and related EDQNM (eddy-damped quasi-normal Markovian) include approximations for such transfer functions.

Energy transfer also occurs in other systems such as molecular collisions and blackbody radiation (from surface to photons), but the complex details of the energetic interactions can be circumvented through the use of the maximum entropy principle (the Second Law), e.g. Maxwell and Planck distribution functions [8, 10]. If left to their own devices, energetic particles or turbulent eddies in this case will organize themselves in the order that maximizes entropy. Similar derivation is possible for the turbulence energy (or power) spectra [11]. This route does not necessitate complex eddy interaction terms since the distribution is representative of the final statistical state. This approach is also referred to as the method of most probable distribution [19], and leads to a universally observable behavior of energetic systems. Other constraints, such as the cut-off at the energy-generating length scale and width of the spectrum as a function of the Reynolds number, can be asserted into the distribution function, leading to a lognormal form (Eq. 1).

The utilitarian aspect of this lognormal-type distribution is evident through the intuitive scaling behavior of spectral parameters, and explicability of the $k^m$ segments in the spectra. In the ascending portion, m = 2 to 4 have been suggested [20], while m=-5/3, -3 and similar have been reported depending on the dimensionality of the turbulent flow. Lognormal functions trace a parabola in log-log graphs, and any one of these tangents can be observed depending on the scale range. With the parametric scaling discussed in this work, full energy spectra can be graphed based on a small number of input parameters such as the Reynolds number ($C_2 \sim 1/Re^m$), viscosity, and energy-producing length and energy scales, in homogeneous and also in some inhomogeneous turbulent flows. In place of limited range of applicability of $k^{-n}$-type of scaling, current theoretical result agrees quite well with the observed energy spectra over the entire range of length and energy scales. For large Reynolds numbers, this means energy scale spanning 8 to 12 orders of magnitude and length scale over 4 to 6. Moreover, the scaling of this spectral equation is very intuitive: when the Reynolds number increases, the range of length scales increases leading to widening of the spectra through $C_2$ term in Eq. 1. Furthermore, as $C_2$ increases the width of the spectra it also raises the magnitude (the energy scale); thus, there is some built-in increase of the energy when the spectra widens at high Reynolds numbers. The peak energy level, however, is determined by other factors such as the total energy generation rate, typically estimated as $U^3/L$. For different turbulence geometry, the peak energy level therefore needs to be adjusted with the factor, $C_1$, in Eq. 1. Likewise, the bending of the spectra at high wavenumbers due to viscosity is easily replicated in the current form. These attributes of the current spectral equation are worth considerations and should prove to be useful.

**REFERENCES**

[1] Kolmogorov, N., A refinement of previous hypotheses concerning the local structure of turbulence in a viscous incompressible fluid at high Reynolds number. J. Fluid Mech., 1962, 13, 82-85.

[2] Kraichnan, R.H., The structure of isotropic turbulence at very high Reynolds numbers. J. Fluid Mech. 1959, 5, 497-543.

[3] Salmon, R., Entropy budget and coherent structures associated with a spectral closure model of turbulence, Journal of Fluid Mechanics, 2018, Vol. 857, pp. 806-822.

[4] Orszag, S., 1970, Analytical theories of turbulence, J. Fluid Mech., 1970, 41, 363.

[5] Boffetta, G. and Ecke, R.E., Two-dimensional turbulence, Annual Review of Fluid Mechanics, 2012, Vol. 44, pp. 427-451.

[6] Brown, T.M., Journal of Physics A, Information theory and the spectrum of isotropic turbulence, 1982, 15, 2285.

[7] Verkley, W.T.M. and Lynch, R., Energy and enstrophy spectra of geostrophic turbulent flows derived from a maximum entropy principle, 2009, Vol. 66, pp. 2216-2236.

[8] Planck, M., Distribution of energy in the spectrum, Ann. Physics, 1901, 4, 3, pp. 553-560.

[9] Cover, T. and Thomas, J., Elements of Information Theory, John Wiley & Sons, Inc., 1991.

[10] Bevensee, R.M., Maximum entropy solutions to scientific problems, Prentice Hall, 1993.

[11] Lee, T.-W., Lognormality in turbulence energy spectra, Entropy, 2020, 22(6), 669.

[12] Comte-Bellot, G. and Corrsin, S., Simple Eulerian time correlation of full- and narrow-band velocity signals in grid-generated isotropic turbulence, J. Fluid Mech., 1971, 48, 2, pp. 273-337.

[13] Champagne, F.H., Friehe, C.A., La Rue, J.C. and Wyngaard, J.C., Flux measurements and fine-scale turbulent measurement in the surface layer over land, J. Atm. Sci., 1977, 34, 515-530.

[14] Saddoughi, S.G. and Veeravalli, S.V., Local isotropy in turbulent boundary layers at high Reynolds numbers, J. Fluid Mech., 1994, 268, pp. 333-372.

[15] Tieleman, H.W., Viscous region of turbulent boundary layer. Colorado State Univ. Rep., 1967, CER 67-68 HWT21.

[16] Kang, H., Stuart C., and Meneveau, C., "Decaying turbulence in an active-grid-generated flow and comparisons with large-eddy simulation." Journal of Fluid Mechanics, 2003, 480, pp. 129-160.

[17] Moser, Robert D., John Kim, and Nagi N. Mansour, Direct numerical simulation of turbulent channel flow up to $Re_\tau$= 590, Physics of Fluids, 1999, 11,4, pp. 943-945.

[18] Tung, K.K. and Orlando, W.W., "The k-3 and k-5/3 energy spectrum of atmospheric turbulence: quasi-geostrophic two-level model simulation", Journal of the Atmospheric Science, 2003, pp. 824-832.

[19] Conrad, K., online notes, https://kconrad.math.uconn.edu/blurbs/analysis/entropypost.pdf

[20] Hinze, J.O., Turbulence, McGraw-Hill Series in Mechanical Engineering. McGraw-Hill, New York, 1975.

[21] Frisch, U., Turbulence, Cambridge University Press, 1995.

[22] Uberoi, M.S. and Freymuth, P., Turbulence energy balance and spectra of the axisymmetric wake, Physics of Fluids, 1970, 13, 2205.

**APPENDIX: Derivation of the Maximum-Entropy Turbulence Energy Spectral Function**

The energy distribution that maximizes the Shannon's entropy under the physical constraints can be obtained using the Lagrange multiplier method [9, 10]. Here, the principal constraint is that the turbulence conserves energy: the kinetic energy is dissipated by viscosity effect progressively at large wavenumbers [20, 21].

$$u'^2 + \nu k^2 u'^2 \delta t = e_o = constant \qquad (A1)$$

u'(k) the turbulent fluctuation velocity at a given wavenumber, k, while $\nu$ is the kinematic viscosity and $\delta$t some time interval. Eq. A1 states that turbulence energy density (on a unit-volume basis) integrated over some time interval $\delta$t is conserved. The above constraint can be transposed into the energy distribution using the Lagrange multiplier method [9]. The first step is to write the objective function F so that

$$F = u'^2 + vk^2 u'^2 \delta t - e_o \tag{A2}$$

The most probable distribution function is found by maximizing logF, following the concept of Shannon's entropy, S=FlogF [9]. Using the Lagrange multiplier method, this distribution has an inverse exponential form [9].

$$E(k)dV = C_1 exp\{-C_2 u'^2 - C_3 k^2 u'^2\}dV \tag{A3}$$

$C_1$, $C_2$ and $C_3$ (=$C_2$⊙) are so-called Lagrange multipliers, to be determined from other constraints. For example, $C_1$ is determined by integrating E(k) to equal the total energy content in the distribution. Converting dV=d($k^{-3}$) to dk basis, we obtain the following energy distribution.

$$E(k) = \frac{C_1}{k^4} exp\{-C_2 u'^2 - C_3 k^2 u'^2\} \tag{A4}$$

In Eq. A4, constants $C_1$, $C_2$, and $C_3$ are determined from the constraints of the turbulence energy content, limiting length scales, and viscosity, respectively. The limiting length scales are the Kolmogorov dissipation length scale and the maximum length scale that exists in the flow. We still need the kinematic scaling for u'(k) in Eq. 4. In Kolmogorov theory [20], u' ~ $k^{-1/3}$ is obtained in the inertial subrange. However, in the current maximum entropy formalism this is an unknown element or a lack of a piece of information. The maximum entropy principle gives the most probable energy distribution under the given physical constraints, but it does not produce unknown information. Thus, the missing pieces of information need to be supplied from observational data, and Eq. A4 provides a framework for testing various kinematic scaling for u'(k). Comparison with observational data can then be used to deduce the empirical form for u'(k) ~ (m-log(k)). We can compare this scaling with u' ~ $k^{-1/3}$ below. Depending on the wavenumber range and scaling constants, there can exist similar slope. Also, Figure A2 shows a comparison of E(k) generated from various kinematic scaling for u'(k), and inverse logarithmic expression gives the best result when compared with experimental data [22]. Given the universality of Eq. A4, as shown in this manuscript, the inverse logarithmic scaling for u' appears most plausible, subject to further experimental verifications.

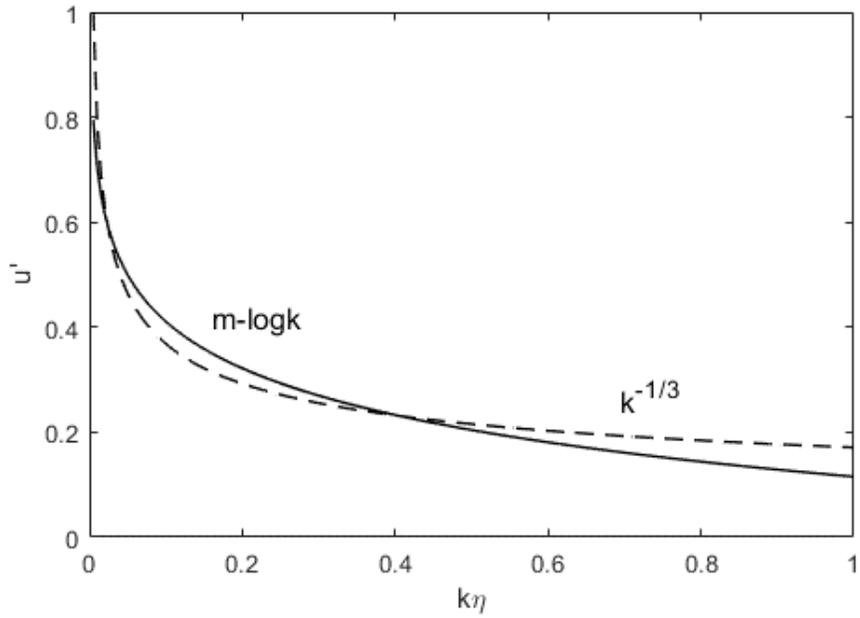

**Figure A1.** Comparison of u'(k)=m-log(k) with $k^{-1/3}$ scaling.

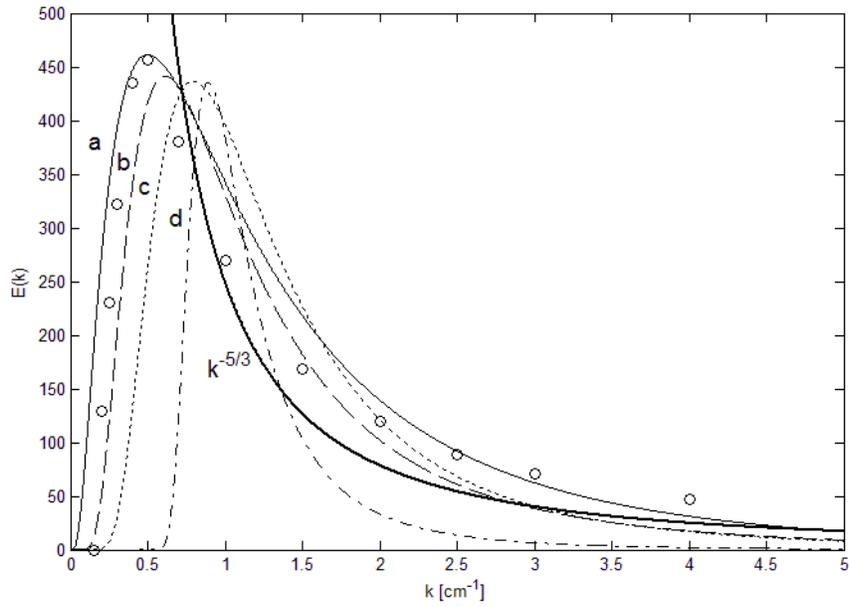

**Figure A2.** Various u'(k) scaling used in Eq. A4 to generate the turbulence energy spectra: a. u'(k)=m-log(k); b. $k^{-1/2}$; c. $k^{-1/3}$; d. $k^{-3}$. Bold line is the Kolmogorov's $k^{-5/3}$ law in the inertial subrange. Symbols are data [22] at $Re_\lambda = 56$. Notice the log-normal shape of the data when graphed in a linear plot.